\documentclass[conference]{IEEEtran}
\IEEEoverridecommandlockouts
\usepackage{cite}
\usepackage{dcolumn}
\usepackage{bm}
\usepackage{bbm}
\usepackage{braket}
\usepackage{hyperref}
\usepackage{physics}
\usepackage{amsmath,amssymb,amsfonts}
\usepackage{algpseudocode}
\usepackage{graphicx}
\usepackage{textcomp}
\usepackage{xcolor}
\usepackage{algorithm}
\usepackage{algpseudocode}
\usepackage{amsmath}
\newtheorem{theorem}{Theorem}

\def\squareforqed{\hbox{\rlap{$\sqcap$}$\sqcup$}}
\def\qed{\ifmmode\squareforqed\else{\unskip\nobreak\hfil
\penalty50\hskip1em\null\nobreak\hfil\squareforqed
\parfillskip=0pt\finalhyphendemerits=0\endgraf}\fi}

\newenvironment{proofof}[1]{\begin{trivlist}\item[]{\flushleft\bf 
Proof of~#1 }}
{\qed\end{trivlist}}

\begin{document}

\title{Nonlocality Distillation Can Outperform Entanglement Distillation\\
}

\author{\IEEEauthorblockN{Peter Høyer}
\IEEEauthorblockA{\textit{Department of Computer Science} \\
\textit{University of Calgary}\\
Calgary, Canada \\
hoyer@ucalgary.ca}
\and
\IEEEauthorblockN{Jibran Rashid}
\IEEEauthorblockA{\textit{Department of Computer Science} \\
\textit{Institute of Business Administration}\\
Karachi, Pakistan \\
jrashid@iba.edu.pk}
\and
\IEEEauthorblockN{Razeen Ud Din}
\IEEEauthorblockA{\textit{Department of Mathematical Sciences} \\
\textit{Institute of Business Administration}\\
Karachi, Pakistan \\
ruddin@iba.edu.pk}}

\maketitle

\begin{abstract}
Given the goal of maximizing CHSH violation, we compare the optimal strategies of entanglement and nonlocality distillation. In the limit of the number of copies of the shared state, entanglement distillation is guaranteed to work by generating a Bell state. For a small number of copies of the state, we show that nonlocality distillation can achieve a higher CHSH value, even though optimal entanglement distillation requires communication. Nonlocality distillation not only outperforms entanglement distillation but also demonstrates superior resource efficiency across multiple metrics for quantum resource estimation. 
\end{abstract}


\section{Introduction}
Protocols that distill resources such as entanglement, correlations and channels perform an important task in quantum information. They allow for the concentration of multiple copies of a noisy resource into a \emph{purer or stronger} version of the resource. A natural consideration is to determine how efficiently the stronger resource can be generated. Here we consider the possibility to distill nonlocal correlations via entanglement distillation and explore whether the resulting bounds are always better than directly performing nonlocality distillation.

Entanglement and nonlocality are distinct but related resources in quantum information theory\cite{Gisin05}. Entanglement refers to the property of composite quantum systems that cannot be described independently, while nonlocality refers to correlations that entangled systems exhibit that cannot be generated classically. Both are crucial resources for quantum communication and computation. However, not all entangled states exhibit nonlocality~\cite{Brunner}. Some entangled states admit a local hidden variable model and thus do not demonstrate nonlocality. This distinction highlights that while nonlocality is a manifestation of entanglement, it is not the same phenomenon.

Entanglement distillation protocols~\cite{Benn92} in quantum information theory allow the extraction of maximally entangled Bell states from multiple copies of weakly entangled or noisy states. Given the importance of Bell states as a unit of entanglement, these protocols are crucial in quantum communication tasks such as quantum teleportation and quantum key distribution (QKD), where high-quality entanglement is essential for secure and reliable information transfer. The fact that some entangled states cannot be distilled (bound entanglement) highlights fundamental limitations in resource conversion~\cite{Bennett1996}. Lo and Popescu~\cite{Lo01} construct the optimal entanglement distillation protocol using one-way communication for bipartite pure states.
 
Peres~\cite{Peres96} initiated the study of nonlocality distillation by identifying collective measurements on multiple copies of a state that can lead to CHSH violation~\cite{CHSH69}, even though a single copy only generates local correlations. Liang and Doherty~\cite{Liang06} provide CHSH violation bounds through collective measurements for $n$ copies of two-qubit pure states. 

Assuming access to only a few copies of a noisy entangled state, we ask the question whether an optimal entanglement distillation protocol also serves as an optimal nonlocality distillation protocol. In the commonly considered framework for entanglement distillation, it is assumed that many identical copies of a noisy entangled state are readily available. If the number of copies is large enough, multiple copies of perfect Bell states can be obtained from the protocol. As a consequence, the issue of comparing the distillation rate of entanglement and nonlocality concentration protocols does not come up. If the number of copies we have available is so small that not even a single perfect Bell state can be deterministically obtained, then it is not known which of the two; entanglement or nonlocality distillation, maximizes CHSH~\cite{CHSH69} violation. We show here that even though entanglement distillation requires communication, for small enough number of copies $n$ of the state and large enough noise, nonlocality distillation can attain a higher CHSH value.

\subsection{Main Results}\label{sec:res}

Our key contribution is to prove optimality of certain existing nonlocality distillation protocols and then compare their performance against optimal protocols for entanglement distillation. For pure states, while the best known protocols are not proven to be optimal, Theorem~\ref{thm:pure} shows that they can already outpeform entanglement distillation.
\begin{theorem}
\label{thm:pure}
Nonlocality distillation for bipartite pure states can attain a higher CHSH value for $n=2$ and $3$ than the optimal entanglement distillation protocol.
\end{theorem}
For mixed states, we are not aware of optimal entanglement distillation protocols for a few copies. We prove in Theorem~\ref{thm:mix} that the best nonlocality distillation protocol for a given mixed state can achieve the same value as the pure state protocol. As a consequence, nonlocality distillation can also perform better for mixed states.
\begin{theorem}
\label{thm:mix}
The optimal nonlocality distillation protocol for mixed states can attain a higher CHSH value for $n=2$ than the optimal entanglement distillation protocol. A bound for $n=3$ copies of the state in Equation~\ref{Eq-1} is given by 
\begin{equation*}
       V \leq 2 \sqrt{
            1 + \left( 1 - 24p^{2} + 80p^{3} - 120p^{4} + 96p^{5} - 32p^{6} \right). 
               }
\end{equation*}
\end{theorem}
In the next Section we formulate the different distillation frameworks and prove Theorems~\ref{thm:pure} and~\ref{thm:mix} in Section~\ref{sec:proof}.

\section{Distillation Review}
\label{sec:rev}
We start by briefly reviewing the optimal distillation protocol for pure states due to Lo and Popescu~\cite{Lo01}.

\subsection{Optimal Entanglement Distillation for Pure States}\label{EC}

Let the states $\ket{\psi}$ and $\ket{\phi}$ be the Bell states given by
\begin{equation*}
\ket{\psi} = \frac{1}{\sqrt{2}}\left(\ket{00} + \ket{11} \right) \quad \text{and} \quad \ket{\phi} = \frac{1}{\sqrt{2}}\left(\ket{01} + \ket{10}\right).
\end{equation*}
We have Alice and Bob in their spatially separated labs share $n$ copies of the pure state  
\begin{equation}
\ket{\Psi}= \sqrt{p}\ket{\psi}+\sqrt{1-p}\ket{\phi}.
\label{eq:state1}
\end{equation}
The Schmidt decomposition of the state $\ket{\Psi}$ is given by 
\begin{equation*}
\ket{\Psi} = 
\left(\frac{\sqrt{p}+\sqrt{1-p}}{\sqrt{2}}\right)\ket{00}
+ \left(\frac{\sqrt{p}-\sqrt{1-p}}{\sqrt{2}}\right)\ket{11}.
\end{equation*}
For $n=3$, the Schmidt decomposition of $\ket{\Psi}^{\otimes n}$ can be stated as
\begin{align*}
\ket{\Phi}_{AB}  &= \hphantom{+}\sqrt{\lambda_{1}} \ket{000}_{A} \ket{000}_{B} + \sqrt{\lambda_{2}} \ket{001}_{A} \ket{001}_{B}  \\
&\quad + \sqrt{\lambda_{3}} \ket{010}_{A} \ket{010}_{B} + \sqrt{\lambda_{4}} \ket{100}_{A} \ket{100}_{B} \\
&\quad + \sqrt{\lambda_{5}} \ket{011}_{A} \ket{011}_{B} + \sqrt{\lambda_{6}} \ket{101}_{A} \ket{101}_{B}  \\
&\quad + \sqrt{\lambda_{7}} \ket{110}_{A} \ket{110}_{B} + \sqrt{\lambda_{8}} \ket{111}_{A} \ket{111}_{B},
\end{align*}
where the coefficients $\lambda_i$ are
\begin{align*}
\lambda_1 &= \left( \frac{\sqrt{p} + \sqrt{1-p}}{\sqrt{2}} \right)^6, \, \lambda_8 = \left( \frac{\sqrt{p} - \sqrt{1-p}}{\sqrt{2}} \right)^6, \\
\lambda_2 &= \lambda_3 = \lambda_4 = \left( \frac{\sqrt{p} + \sqrt{1-p}}{\sqrt{2}} \right)^4 \left( \frac{\sqrt{p} - \sqrt{1-p}}{\sqrt{2}} \right)^2, \\
\lambda_5 &= \lambda_6 = \lambda_7 = \left( \frac{\sqrt{p} + \sqrt{1-p}}{\sqrt{2}} \right)^2 \left( \frac{\sqrt{p} - \sqrt{1-p}}{\sqrt{2}} \right)^4,
\end{align*}
with $\lambda_i \geq \lambda_{i+1}$ and $\sum_{i=1}^{8} \lambda_i = 1$. Here, the subscripts $A$ and $B$ refer to Alice and Bob's share of the state $\ket{\Phi}_{AB}$ respectively. The optimal probability $p_{\textrm{succ}}$ to successfully distill the Bell state $\ket{\psi}$ from $\ket{\Phi}_{AB}$ is determined by Equation~\ref{prob}. 
\begin{figure*}
{\small
\begin{equation}\label{prob}
p_{\textrm{succ}} = \min_{1 \leq r \leq 2} \frac{2}{r} \left( \lambda_{2-r+1} + \lambda_{2-r+2} + \cdots + \lambda_8 \right)=
 \min \left\{ 1,6\left( \frac{\sqrt{p}+\sqrt{1-p}}{\sqrt{2}}\right)^2 \left( \frac{\sqrt{p} - \sqrt{1-p}}{\sqrt{2}} \right)^2+2\left( \frac{\sqrt{p} - \sqrt{1-p}}{\sqrt{2}} \right)^6 \right\}
\end{equation}    
}
\end{figure*}
If the distillation protocol fails, the resulting state is a product state. This allows us to determine the CHSH value $V_{ED}$ obtained for the $3$ copies as a weighted average
\begin{equation*}
V_{ED} 
= 2\sqrt{2}p_{\textrm{succ}} + 2(1 - p_{\textrm{succ}}).
\end{equation*}

\subsection{Nonlocality Distillation for Pure States}\label{NLD}

Much of the recent work on nonlocality distillation has focused on the box world framework~\cite{Swolf08a},~\cite{Allcock09a},~\cite{Hoyer10},~\cite{Forster11}. For $n$ copies of the pure state $\ket{\Psi}$ in Equation~\ref{eq:state1} the best known CHSH bound $V_{ND}$ for nonlocality distillation due to Liang and Doherty~\cite{Liang06} is
\begin{equation}
V_{ND} = 2 \sum_{n=1}^{\lfloor{d/2}\rfloor} \sqrt{\left(\lambda_{2 n-1}^2+\lambda_{2 n}^2\right)^2+4 \lambda_{2 n}^2 \lambda_{2 n-1}^2} ,
\label{eq:pureval}
\end{equation}
where $\lambda_{i}$ are Schimdt coefficients and $d$ is the dimension of the state. For $n=1$ the bound is known to be optimal~\cite{gisin1992maximal}, but optimality is not proven for $n>1$. For $n=1$, the observables that achieve the bound are given by 
\begin{align}
\begin{split}
 A_{0}&=Z\\
 A_{1}&=X \\
 B_{0}&=\frac{{(1-2p)}}{\sqrt{1+(1-2p)^2}}Z+\frac{1}{\sqrt{1+(1-2p)^2}}X \\
 B_{1}&=\frac{{(1-2p)}}{\sqrt{1+(1-2p)^2}}Z-\frac{1}{\sqrt{1+(1-2p)^2}}X.    
\end{split}
\label{eq:meas}
\end{align}
For these measurements, the four CHSH correlators take the values
\begin{align*}
\bra{\Psi}A_0 \otimes B_0 \ket{\Psi} &=\hphantom{-} \bra{\Psi}A_0 \otimes B_1 \ket{\Psi} =\frac{(1 - 2p)^2}{\sqrt{2 - 4p + 4p^2}}\\
\bra{\Psi}A_1 \otimes B_0 \ket{\Psi} &= -\bra{\Psi}A_1 \otimes B_1 \ket{\Psi} = \frac{1}{\sqrt{2 - 4p + 4p^2}}
\end{align*}
which results in the CHSH bound
\begin{equation} \label{eq:bound}
  V = 2\sqrt{(1+(1-2p)^2)} .
\end{equation}
For $n=3$ the nonlocality bound is
\begin{equation*}
V_{ND} = 2 \sum_{n=1}^{4} \sqrt{\left(\lambda_{2 n-1}^2+\lambda_{2 n}^2\right)^2+4 \lambda_{2 n}^2 \lambda_{2 n-1}^2} ,
\label{eq:pureval3}
\end{equation*}
however, the corresponding measurements have only been verified numerically. 

\subsection{Nonlocality Distillation for Mixed States}\label{sec-05}

Given that we do not have a canonical representation for mixed states, motivated by our pure state analysis, we consider the mixed state
\begin{equation}\label{Eq-1}
\rho = p\ket{\psi}\bra{\psi}+(1-p)\ket{\phi}\bra{\phi}
\end{equation}
where,
The CHSH value is then given by
\begin{equation*}
\Tr(M \rho) = \Tr((pM + (1-p)M') \ket{\psi}\bra{\psi})
\end{equation*}
where,
\begin{align*}
 M &= A_0 \otimes B_0 + A_0 \otimes B_1 + A_1 \otimes B_0 - A_1 \otimes B_1 \\
M' &= A_0 \otimes B_0'  + A_0 \otimes B_1' + A_1 \otimes  B_0' - A_1 \otimes B_1' 
\end{align*}
and $B_i' = X B_i X$. Let
\begin{equation} \label{eq:N}
   N = pM+(1-p) M'
\end{equation}
Using the same observables as given in Equation~\ref{eq:meas} we obtain
\begin{align*}
\Tr(A_0 \otimes B_0 \rho) &=\hphantom{-} \Tr(A_0 \otimes B_1 \rho) =\frac{(1 - 2p)^2}{\sqrt{2 - 4p + 4p^2}}\\
\Tr(A_1 \otimes B_0 \rho) &= -\Tr(A_1 \otimes B_1 \rho) = \frac{1}{\sqrt{2 - 4p + 4p^2}},
\end{align*}
which results in the same CHSH bound as the pure state, i.e.,
\begin{equation*}
V= 2\sqrt{1+(1-2p)^2}.
\end{equation*}
We prove in Theorem~\ref{thm:mix} that the bound for two copies of $\rho$ is the same as for $n=1$. Note that a similar property seems to hold for the pure state case as well, even though its not proven.

For $n=3$ copies of state $\rho$, we prove the following bound in Section~\ref{sec:proof}
\begin{equation*}\label{Bound:mix3}
    2 \sqrt{
            1 + \left( 1 - 24p^{2} + 80p^{3} - 120p^{4} + 96p^{5} - 32p^{6} \right) 
               },
\end{equation*}
but note it is not known to be tight. Figure~\ref{fig06} shows the bounds for one, two and three copies of the mixed state.

\section{Comparison of Entanglement \& Nonlocality Distillation} \label{sec:proof}

\begin{figure*}[htbp]
    \centering
    \includegraphics[width=\linewidth]{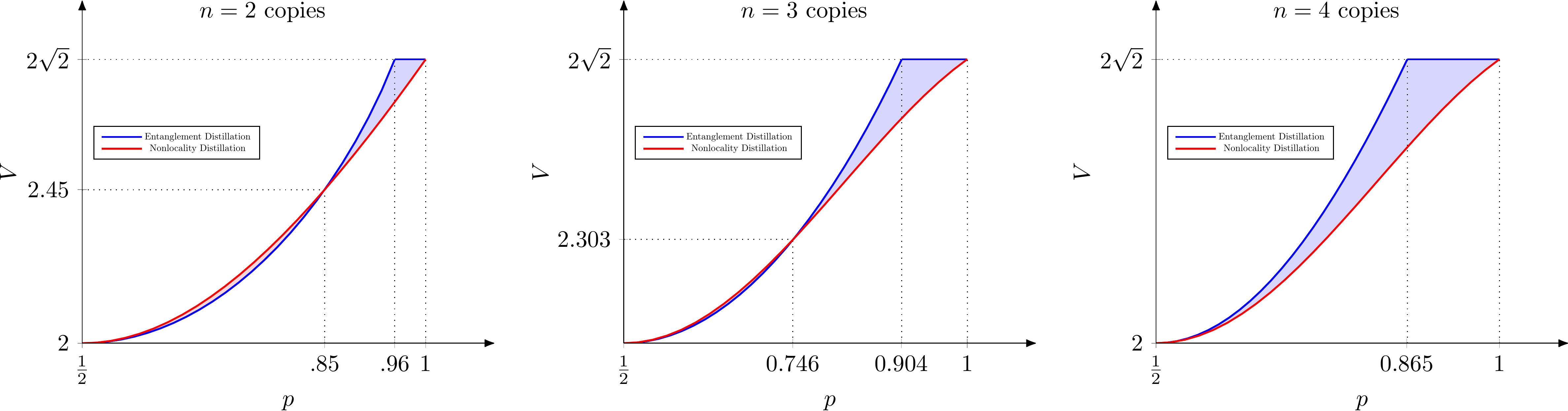}
    \caption{Comparison between entanglement distillation and nonlocality distillation for $n=2, 3$ and $4$ for pure states. The results indicate that nonlocality distillation outperforms entanglement distillation for a specific range of $p$, but this advantage vanishes beyond three copies.}
    \label{combine_plot}
\end{figure*}

The values attained in Theorem~\ref{thm:pure}, which we now prove, can be visualized in Figure~\ref{combine_plot}. 
\begin{proofof}{Theorem~\ref{thm:pure}}
Using the entanglement and nonlocality distillation protocols from Sections~\ref{EC} and~\ref{NLD} respectively, for $n=2$ we obtain
\begin{align*}
    V_{ED}&=2+4(2\sqrt{2}-2)t+2\left( \frac{\sqrt{p} + \sqrt{1-p}}{\sqrt{2}} \right)^4\\
    &\leq 2\sqrt{1+4t} = V_{ND} \text{ for } p \in [0.5,0.85],
\end{align*}
where 
\begin{equation*}
t=\left( \frac{\sqrt{p} + \sqrt{1-p}}{\sqrt{2}} \right)^2\left( \frac{\sqrt{p} - \sqrt{1-p}}{\sqrt{2}} \right)^2.
\end{equation*}
This shows that for the interval $p \in [0.5,0.85]$, nonlocality distillation achieves a higher value than entanglement distillation. Figure~\ref{combine_plot} (left) shows this comparison. 

Similarly, for $n=3$ we have
\begin{align*}
V_{ED}&=2+6(2\sqrt{2}-1)t+2\left( \frac{\sqrt{p} - \sqrt{1-p}}{\sqrt{2}} \right)^6\\
&\leq (4\sqrt{2})t+2(\sqrt{1+4t})r = V_{ND} \text{ for } p \in [0.5,0.746],
\end{align*}
where 
\begin{equation*}
r=\left( \frac{\sqrt{p} + \sqrt{1-p}}{\sqrt{2}} \right)^4+\left( \frac{\sqrt{p} - \sqrt{1-p}}{\sqrt{2}} \right)^4.
\end{equation*}
This proves that for $p \in [0.746,0.904]$ nonlocality distillation again outperforms entanglement distillation. Figure~\ref{combine_plot} (middle) shows this comparison and (right) shows the situation for $n=4$ where entanglement distillation is always better.
\end{proofof}

\begin{figure}[htbp]
\centerline{\includegraphics{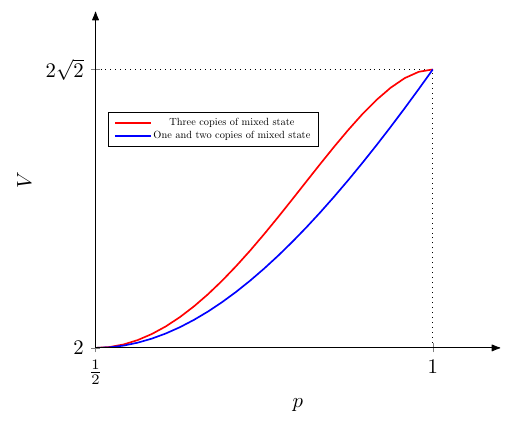}}
\caption{Nonlocality distillation bounds for one, two and three copies of the mixed state $\rho$.}
\label{fig06}
\end{figure}
We now proceed to prove Theorem~\ref{thm:mix} using the  mixed state in Equation~\ref{Eq-1} for distillation. The bound we achieve for $n=1$ and $2$ matches the value in Equation~\ref{eq:bound} for the pure state case. 
\begin{proofof}{Theorem~\ref{thm:mix}}
We note that the square of the operator $N$ in Equation~\ref{eq:N} for $n$ copies of the mixed state is given by
\begin{equation} \label{eq:nsquare}
N^2=4 {\mathbbm{1}} \otimes \tilde{K}+(2 A) \otimes \tilde{L},
\end{equation}
where $A=A_1A_0-A_0A_1$ and
\begin{align}
 \tilde{K} &=\sum_{x,y \in \{0,1\}^n} K[x, y] \cdot \frac{1}{2}\left(B_0^x B_0^y+B_1^x B_1^y\right), \\
 \tilde{L} &=\sum_{x,y \in \{0,1\}^n} L[x, y] \cdot \frac{1}{2}\left(B_0^x B_1^y-B_1^y B_0^x\right).
\end{align}
Here we have $B_{0}^{x} = X^x B_0 X^x$ and $B_{1}^{x} = X^x B_1 X^x$ with $X^x = \otimes_{i=1}^n X^{x_i}$. In general, we also have $K=L$ with the following explicit forms for $n=1,2$ and $3$.

\begin{enumerate}
    \item For $n=1$, we have
\begin{equation*}
    K = L =
\begin{bmatrix}
a & b \\
b & a
\end{bmatrix}.
\end{equation*}

where we fix

\begin{align*}
a &= \frac{p^2+(1-p)^2}{2} \textrm{ and} \\
b &=p(1-p).
\end{align*}

\item For $n=2$, we have
\begin{equation*}
K=L =
\begin{bmatrix}
a & b & b & c \\
b & a & c & b \\
b & c & a & b \\
c & b & b & a
\end{bmatrix},
\end{equation*}
where we fix
\begin{align*}
a &= \frac{(p^2+(1-p)^2)^2}{4}, \\
b &= \frac{p(1-p)(p^2+(1-p)^2)}{2} \textrm{ and} \\ 
c &=p^2(1-p)^2.
\end{align*}

\item For $n=3$, we have
\begin{equation*}
K=L =
\begin{bmatrix}
a & b & b & c & b & c & c & d \\
b & a & c & b & c & b & d & c \\
b & c & a & b & c & d & b & c \\
c & b & b & a & d & c & c & b \\
b & c & c & d & a & b & b & c \\
c & b & d & c & b & a & c & b \\
c & d & b & c & b & c & a & b \\
d & c & c & b & c & b & b & a
\end{bmatrix},
\end{equation*}
where we set
\begin{align*}
a &= \frac{(p^2+(1-p)^2)^3}{8}, \\
b &= \frac{p(1-p)((p^2+(1-p)^2))^2}{4}, \\
c &= \frac{p^2(1-p)^2(p^2+(1-p)^2)}{2}\textrm{ and} \\ 
d&=p^3(1-p)^3.
\end{align*}
\end{enumerate}
We have $\norm{ \tilde{K} } \leq 1$ , $\norm{ 4 {\mathbbm{1}} } = 4$ and $\norm{ 2A } \leq 4$. In order to bound $\norm{ \tilde{L} }$, let $C_{xy} = B_0^x B_1^y-B_1^y B_0^x$ so, for $n=1$ we have
\begin{align*}
    \tilde{L} =&\sum_{x,y \in \{0,1\}} L[x, y] \cdot \frac{1}{2}C_{xy}\\
     =& \hphantom{+} \frac{a}{2}C_{00}+ \frac{b}{2}C_{01}+
\frac{b}{2}C_{10}+\frac{a}{2}C_{11}.
\end{align*}
We can restate the expression as
\begin{equation*}
\tilde{L}=\hphantom{+} \frac{a}{2}C_{00}- \frac{b}{2}C_{00}-
\frac{b}{2}C_{11 }+\frac{a}{2}C_{11} + \frac{b}{2}\sum_{x,y \in \{0,1\}}C_{xy}.
\end{equation*}
Using the cyclic property we have
\begin{align*}
B_0^0 B_1^1-B_1^1 B_0^0 &= B_0XB_1X-XB_1XB_0\\
&=XB_1XB_0-XB_1XB_0 = 0
\end{align*}
and similarly $B_0^1 B_1^0-B_1^0 B_0^1=0$. We note that the summation term $\norm{\frac{b}{2}\sum_{x,y \in \{0,1\}}C_{xy}}$ vanishes and we obtain the bound
\begin{equation*}
    \norm{\tilde{L}} \leq 2|a-b|=  (1-2p)^2.
\end{equation*}
Finally, combining all the pieces we have
\begin{align*}
    \norm{N^2} &\leq  \norm{{4} {\mathbbm{1}}}  \cdot \norm{\tilde{K}}+\norm{2 A} \cdot \norm{\tilde{L}}\\
    & \leq 4 + 4(1-2p)^2,
\end{align*}
which implies as required
\begin{equation*}    
\norm{N}  \le 2\sqrt{(1+(1-2p)^2)}.
\end{equation*}
In the case $n=2$, we similarly bound $\tilde{L}$ by noting that
\begin{align*}
    \tilde{L} =&\sum_{x,y \in \{0,1\}^2} L[x, y] \cdot \frac{1}{2}C_{xy}\\
     =& \hphantom{+} \frac{a-b}{2}\big(C_{0000}+C_{0011}+C_{0101}+C_{0110}\big)\\
     &+\frac{c-b}{2}\big(C_{0000}+C_{0011}+C_{0101}+C_{0110}\big) \\
     &+ \frac{b}{2}\sum_{x,y \in \{0,1\}^2}C_{xy},
\end{align*}
which gives
\begin{equation*}
    \norm{\tilde{L}} \leq 4|a-b|+4|c-b| =  (1-2p)^2.    
\end{equation*}
This gives a bound equal to the $n=1$ case,~i.e.,
\begin{equation*}    
\norm{N}  \leq 2\sqrt{(1+(1-2p)^2)}.
\end{equation*}
For $n=3$, we may express $\tilde{L}$ as
\begin{equation*}
    \tilde{L} =L_1+ \frac{b}{2}\sum_{x,y \in \{0,1\}^3}C_{xy},
\end{equation*}
where $L_1$ is given by
\begin{align*}
&\resizebox{\columnwidth}{!}{$
\begin{bmatrix}
a-b & 0 & 0 & c-b & 0 & c-b & c-b & d-b \\
0 & a-b & c-b & 0 & c-b & 0 & d-b & c-b \\
0 & c-b & a-b & 0 & c-b & d-b & 0 & c-b \\
c-b & 0 & 0 & a-b & d-b & c-b & c-b & 0 \\
0 & c-b & c-b & d-b & a-b & 0 & 0 & c-b \\
c-b & 0 & d-b & c-b & 0 & a-b & c-b & 0 \\
c-b & d-b & 0 & c-b & 0 & c-b & a-b & 0 \\
d-b & c-b & c-b & 0 & c-b & 0 & 0 & a-b
\end{bmatrix}.
$}
\end{align*}
$\norm{L_1}$ can be bounded by 
\begin{equation*}
    \norm{L_1} \leq 8|a-b|+24|c-b|+8|d-b|,
\end{equation*}
which finally gives
\begin{equation*}    
\norm{N}  \leq 2 \sqrt{
            1 + \left( 1 - 24p^{2} + 80p^{3} - 120p^{4} + 96p^{5} - 32p^{6} \right) 
               }.
\end{equation*}
\end{proofof}
Figure~\ref{fig06} plots the mixed state bounds achieved in Theorem~\ref{thm:mix}. It is open whether the bound for $n=3$ is tight.


\subsection{Quantum Resource Estimation for Distillation}

One factor to consider when comparing different distillation protocols is the resource requirements for their implementation. Determining these resources is even more relevant for NISQ hardware due to limited coherence times and number of available Bell states. Quantum resource estimation~\cite{Hansen24,Quetschlich24,vanDam23} allows us to determine the number of qubits and quantum gates, processing time and other resources needed to run a quantum protocol, assuming specific hardware characteristics. Examples of different quantum resource estimators include pyLIQTR~\cite{pyliqtr}, Qualtran~\cite{Qualtran}, BenchQ~\cite{benchq} and Azure Quantum Resource Estimator~\cite{estimator}.


State preparation is an essential part of quantum computing and can be more complex than its classical counterpart. In fact, it is known that some $n$-qubit states can only be prepared with circuits of size $\Omega(2^n)$~\cite{knill1995approximation}. Gleinig and Hoefler~\cite{gleinig2021efficient} construct an algorithm that takes advantage of sparsity to make sparse state preparation asymptotically more efficient.







\begin{figure}[htbp]
    \centering
    \includegraphics[width=0.45\textwidth]{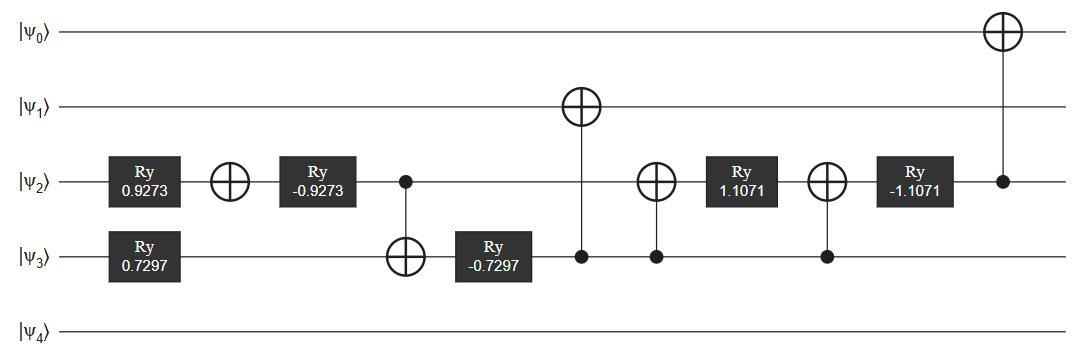}
    \caption{Circuit for preparing two copies of the initial quantum state $\ket{\Psi}$. The fifth qubit is n ancilla for the entanglement distillation process.}
    \label{Sparse_state}
\end{figure}
\begin{figure}[htbp]
    \centering
    \includegraphics[width=0.4\textwidth]{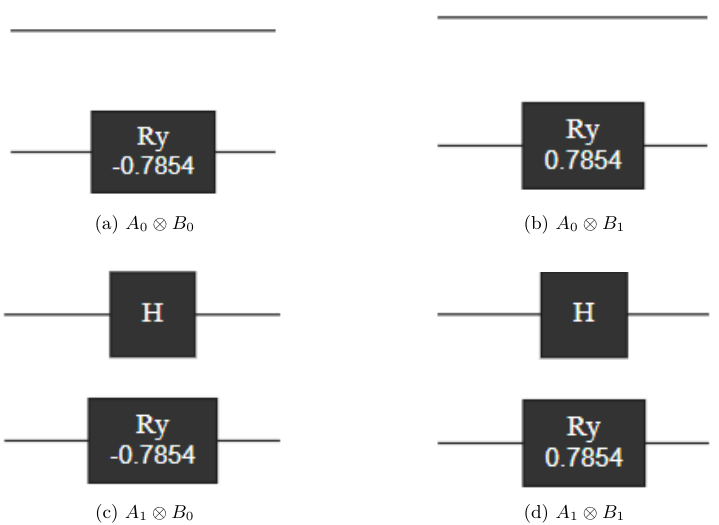}
    \caption{The optimal measurement circuit for entanglement distillation. The optimal CHSH measurements are performed on the first qubit of each party (qubits $0$ and $3$ in Figure~\ref{ed:circ}), since the Bell pair is generated between these qubits.}
    \label{ed:meas}
\end{figure}
\begin{figure}[htbp]
    \centering
    \includegraphics[width=0.4\textwidth]{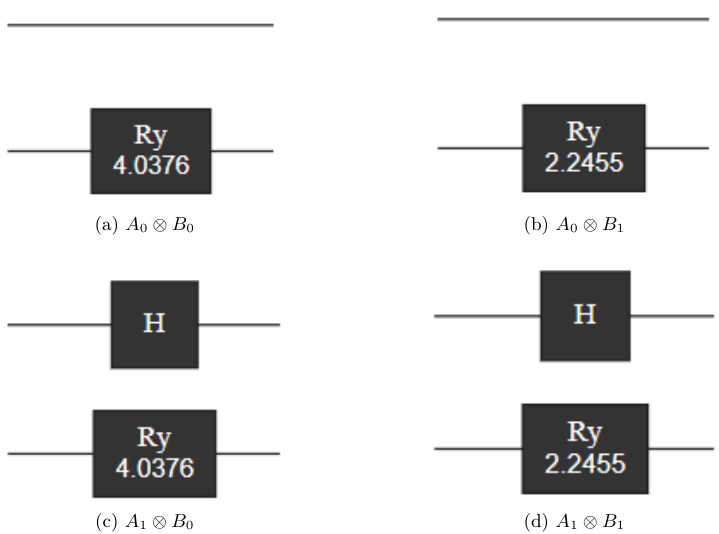}
    \caption{The circuit for optimal nonlocality distillation measurements given by Equation~\ref{eq:meas}.}
    \label{nd:meas}
\end{figure}

Using our implementation of the sparse state algorithm~\cite{Razeen2025QuantumSparse}, we estimate the quantum resources for nonlocality and entanglement distillation for the $n=2$ pure state case. The entire process can be divided into the following three phases.
\begin{equation*}
     \textit{Preparation} \mapsto  \textit{Distillation} \mapsto \textit{Measurement}
\end{equation*}
Although in an actual distillation scenario the states are provided as input, in simulation we need to prepare the states in phase $1$. The circuit in Figure~\ref{Sparse_state} corresponds to the state preparation circuit for both nonlocality and entanglement distillation. The measurements in phase $3$ (Figures~\ref{ed:meas} and ~\ref{nd:meas}) also result in relatively similar circuits for both protocols. 

\begin{figure}[htbp]
    \centering
    \includegraphics[width=0.45\textwidth]{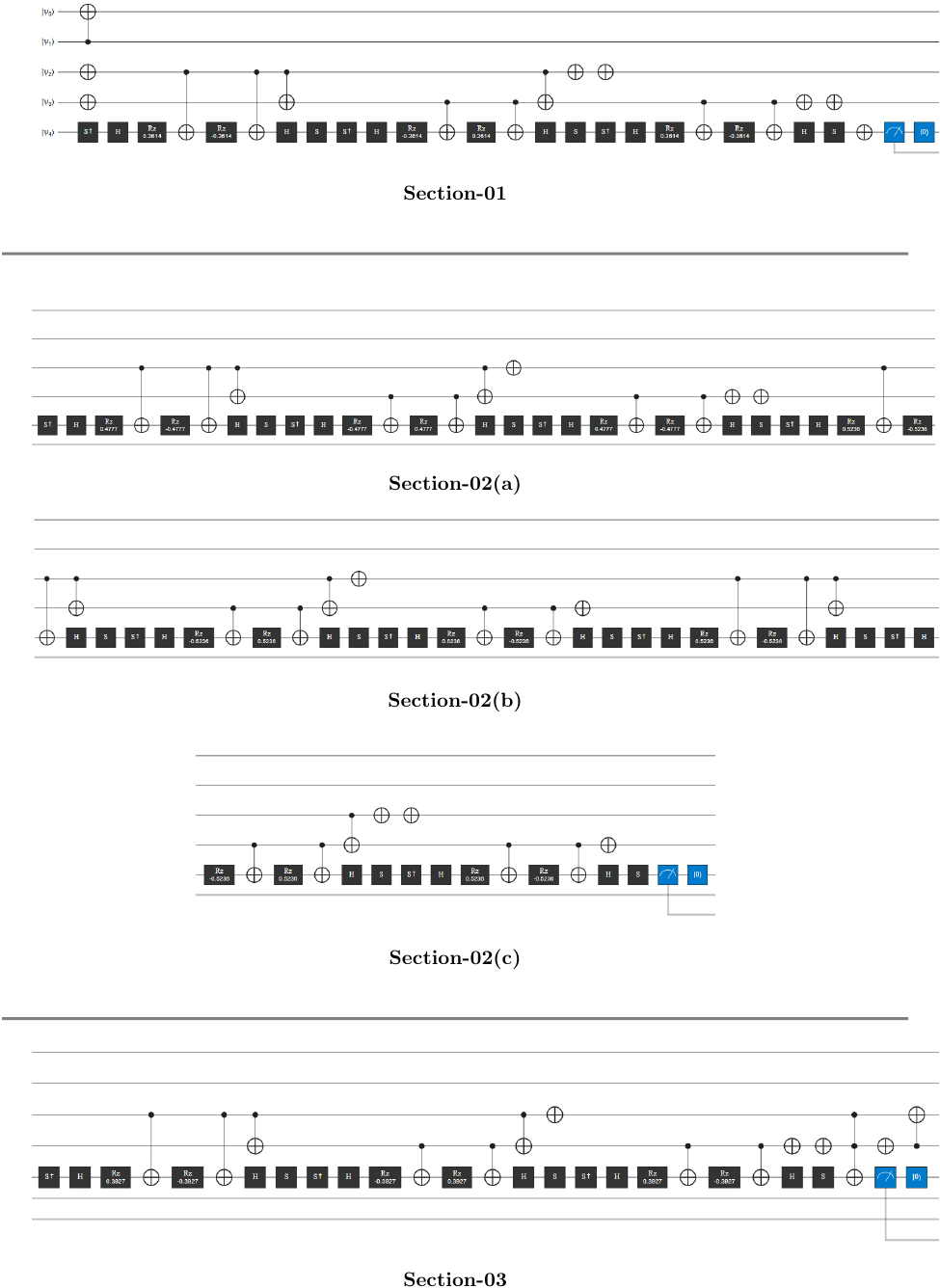}
    \caption{The entanglement distillation circuit that generates a Bell state from two copies of a noisy state. The circuit is divided into three sections, corresponding to three measurements on the ancilla qubit (qubit $4$).
    }
    \label{ed:circ}
\end{figure}

The key distinction occurs in entanglement distillation during phase $2$ (Figure~\ref{ed:circ}). It is the main resource-intensive task during the process; and this is true without even taking into account the need for communication between parties. This is expected since it corresponds to the work needed to distill a quantum state. However, nonlocality distillation does not have any similar task to perform. This  highlights the significantly greater experimental resource requirement for entanglement distillation in comparison to nonlocality distillation.

\begin{table}[htbp]
\caption{Resource requirements for Nonlocality versus Entanglement Distillation for Two copies of $\ket{\Psi}$ with Azure Resource Estimator}
\centering
\small
\renewcommand{\arraystretch}{1.2}

\resizebox{\columnwidth}{!}{%
\begin{tabular}{|c|c|c|c|c|}
\hline
 & \multicolumn{4}{|c|}{\textbf{Resource Metrics}} \\ 
\cline{2-5}
 & \textbf{\textit{Logical Qubits}} & \textbf{\textit{Logical Depth}} & \textbf{\textit{No. of T States}} & \textbf{\textit{Runtime (s)}} \\ 
\hline
Nonlocality Distillation  & 15 & 63 & 150 & 0.00025 \\ 
\hline
Entanglement Distillation  & 18 & 512 & 500 & 0.00225 \\ 
\hline
\end{tabular}%
}
\label{tab:comparison}
\end{table}
Table~\ref{tab:comparison} illustrates the resource requirements for implementing the nonlocality and entanglement distillation protocols using two copies of $\ket{\Psi}$. It is evident from the comparison that nonlocality distillation requires significantly fewer resources. 



\section{Discussion}

It still remains to prove optimality bounds in general for pure and mixed state nonlocality distillation along with their corresponding measurements. One criticism of our comparison is the need for communication during entanglement distillation. If the players can communicate, Alice can just communicate her input to Bob or vice versa and there is no need to distill. However, the bit communicated during entanglement distillation contains zero information about the player's input. It is used exclusively to distill the quantum state. So, given that nonlocality distillation can outperform entanglement distillation even though the latter uses communication, only reinforces the view that entanglement and nonlocality are different resources.

 Recently, Sun et al.~\cite{Sun25} have shown that entanglement distillation can be used to improve distillability of secret keys in quantum key distribution. Is it similarly possible to obtain even a higher key rate by using the better performance of nonlocality distillation? Device independent QKD allows for the successful sharing of key between Alice and Bob even when the physical device used to generate the key is not trusted. 

 Farkas et al.~\cite{Farkas21} have shown that there exist nonlocal correlations obtained by measurements on Werner states that either do not allow for secret key generation or require new reconciliation techniques. So, nonlocal correlations do not automatically imply secrecy for DIQKD protocols. At the same time, it is also known that unbounded DIQKD rates are possible with arbitrarily small nonlocality~\cite{Wooltorton24}. Hence, the secrecy of the key depends on the structure of nonlocal correlations shared between the parties during the generation process. We propose using nonlocality distillation as a possible reconciliation technique during these protocols. Finally, such a technique could provide examples where the limits of tolerable noise for key agreement are different for classical and quantum protocols~\cite{Gis00}.



\bibliographystyle{IEEEtran}
\bibliography{ref}

\end{document}